\documentclass{article}

\usepackage{graphicx}
\graphicspath{{Figures/}}
\usepackage{amssymb}
\usepackage{bbding}
\usepackage[usenames, dvipsnames]{color}
\usepackage[paperwidth=8.5in, paperheight=9.5in]{geometry}
\usepackage{subfig}

\usepackage{multirow}
\usepackage{microtype}
\usepackage{cite}
\usepackage{booktabs}

\usepackage{tabu}

\usepackage{xspace}

\usepackage{verbatim}

\usepackage{array}
\newcolumntype{L}[1]{>{\raggedright\let\newline\\\arraybackslash\hspace{0pt}}m{#1}}
\newcolumntype{C}[1]{>{\centering\let\newline\\\arraybackslash\hspace{0pt}}m{#1}}
\newcolumntype{R}[1]{>{\raggedleft\let\newline\\\arraybackslash\hspace{0pt}}m{#1}}

\title{Towards Understanding Enjoyment and Flow in Information Visualization}
\author{Bahador Saket, Carlos Scheidegger and Stephen Kobourov\\Computer Science Department, University of Arizona}
\date{}

\begin{document}
\maketitle

\begin{abstract}
Traditionally, evaluation studies in information visualization have measured effectiveness by assessing performance time and accuracy. More recently, there has been a concerted effort to understand aspects beyond time and errors. In this paper we study enjoyment, which, while arguably not the primary goal of visualization, has been shown to impact performance and memorability. Different models of enjoyment have been proposed in psychology, education and gaming; yet
there is no standard approach to evaluate and measure enjoyment in visualization. 
In this paper we relate the flow model of Csikszentmihalyi to Munzner's nested model of visualization evaluation and previous work in the area. We suggest that, even though previous papers tackled individual elements of flow, in order to understand what specifically makes a visualization enjoyable, it might be necessary to measure all specific elements.
\end{abstract}

\section{Introduction} \label{SEC:Introduction}
Within visualization, traditional usability and user experience research has focused on two aspects: measuring performance time and accuracy. 
For example, four out of five papers in the evaluation track published at InfoVis 2014 measured performance time and accuracy to compare different visualization techniques~\cite{IV2014_Borgo, IV2014_Chevalier,IV_Saket, IV2014_Fuchs}. 
Although several recent papers study memorability ~\cite{Hullman_Engagement,IV_RB,Isola2011,IV_Borkin}, other aspects such as enjoyment or engagement are not as developed.

This focus on time and accuracy can be explained in part by how easy it is to tell how quickly or accurately a user performs a task, especially in contrast to aspects that are harder to define, much less measure.
Further, one might argue that providing enjoyment is not a primary goal in visualization~\cite{vov}.
Still, it is likely worthwhile to understand the impact of factors such as enjoyment on performance; for example, positive mental states appear linked to better problem-solving performance in general~\cite{Fredrickson,Isen}, and in InfoVis in particular~\cite{Harrison_priming}.

Bateman et al.~\cite{CHI_SB} have tested the comprehension and recall of plain and embellished charts. During the course of their evaluation, they asked subjects to rate how they enjoyed the process, and their results suggest 
that embellished charts are more enjoyable than plain ones. Li et al.~\cite{IV_li2014} have recently reported a replication of both recall and enjoyment effects of the original study.

Assume for now that the results of these studies hold in general, and so embellished charts are truly more enjoyable than plain ones. We are then confronted with the following scenario: as mindful visualization designers, we would prefer to avoid chartjunk~\cite{Tufte1}. At the same time, we would like to elicit positive mental states in visualization consumers (including enjoyment), for the reasons we mentioned earlier.
The relevant questions are then: what specifically makes some visualizations enjoyable?

One of the most well known model for understanding and measuring enjoyment in psychology is the flow model by Csikszentmihalyi~\cite{Csikszentmihalyi}. He first studied people who put a great deal of time and effort on challenging activities such as rock climbing and chess playing.
The results revealed that various activities were described similarly when they were going well: as Csikszentmihalyi writes, ``the way a long-distance swimmer felt when crossing the English Channel was similar to the way a chess player felt during a tournament, or a climber progressing up a difficult rock face''. 

In visualization, the best known model for understanding evaluations is Munzner's nested model~\cite{IV_Munzner}.
In this paper, we discuss how elements of the flow model correspond to levels of Munzner's model, and how previous work has indirectly measured some of these flow elements.

We begin with a brief summary of the flow model and identify elements that capture the experience of flow. In section 3 we review several studies in other fields which have used flow model 
to either design a new model or assess user enjoyment.
In section 4 we discuss how the nested model relates to specific parts of flow, and, in particular, how one flow element appears fundamentally different from the others, and how this potentially impacts visualization evaluation. We conclude with some possible recommendations for evaluation studies in visualization that try to understand enjoyment.

\begin{table}
 \centering
  \tiny
  \begin{tabular}{|p{0.09\textwidth}|p{0.14\textwidth}|p{0.12\textwidth}|p{0.16\textwidth}|p{0.10\textwidth}|p{0.18\textwidth}|}
    &       Why?       &       What?      &       How?             & How Fast? \\ 
    & problem / domain & data / operation & encoding / interaction & algorithm & references \\
    \hline
    Challenge  & & \checkmark & \checkmark & & Hullman et al.~\cite{Hullman_Engagement} \\
    \hline
    Focus      & & & \checkmark & & --- \\
    \hline
    Clarity    &\checkmark&\checkmark& & & Brehmer and Munzner~\cite{IV_Brehmer} \\
    \hline
    Feedback   & & &\checkmark&\checkmark& Liu and Heer~\cite{Liu:2014:TEO} \\
    \hline
    Control    & & &\checkmark& & Kondo and Collins~\cite{Collins_Dimpvis} \\
    \hline
    Immersion  & & &\checkmark& & van Dam et al.~\cite{vanDam} \\
    \hline
  \end{tabular}
  \caption{In this paper, we relate enjoyment, as encoded by Csikszentmihalyi's flow model (rows) to visualization design and evaluation, through Munzner's nested model (columns). The ``references'' column, while making no attempts at comprehensiveness, shows examples of previous work in visualization which appear to be related to specific elements of the flow model. }
\end{table}

\section{Flow}
Csikszentmihalyi conducted a series of experiments in different countries in which he asked people to explain how and when they achieved the highest level of enjoyment when performing some activity~\cite{Csikszentmihalyi}. 
As Csikszentmihalyi writes, ``Regardless of culture, social class, gender or age, the respondents described enjoyment in very much the same way. \textit{What} they did to experience enjoyment varied dramatically --- the elderly Koreans liked to meditate, the teenage Japanese liked to swarm around in motorcycle gangs --- but they described \textit{how} it felt when they enjoyed themselves in almost identical terms.''~\cite{Csikszentmihalyi} He identifies, among others, the following factors as encompassing the experience of flow:

\begin{itemize}
 \item Challenge: the activity must be challenging and require skill
 \item Focus: it should be possible to concentrate on the task
 \item Clarity: it should be possible to concentrate on the activity \emph{because it has clear goals}
 \item Feedback: it should be possible to concentrate on the activity \emph{because it provides immediate feedback}
 \item Control: participants should feel a sense of control over actions
 \item Immersion: participants should lose the concern for self (this is sometimes described as being ``in the zone'')
\end{itemize}
Very different experiences, when engaging and enjoyable, elicit feelings described in similar ways. We speculate that enjoyable, engaging visualizations should elicit comparable descriptions.

\section{Previous Work}
The flow model has been applied by researchers in other fields to assess enjoyment~\cite{Pachet_flow, Education_Csikszentmihalyi,Flow_Music_Book} and to create new models~\cite{Sweetser_flow, Vass_flow}. 
A multi-year study of student experiences in two different educational settings (Montessori and traditional) found that Montessori settings~\cite{Education_Csikszentmihalyi} helped students to achieve flow experiences more frequently~\cite{Education_Csikszentmihalyi}. 
Vass et al.~\cite{Vass_flow} 
 combined several theories, including the flow model, 
 for the development of problem solving environments that support creativity. 
The flow model has also been used in a framework for constructing engaging commercial websites\cite{Jennings_flow}, to assess enjoyment in an interactive music environment~\cite{Pachet_flow}, and to assess information systems~\cite{Artz_flow}. 

Sweetser et al.~\cite{Sweetser_flow} combine various heuristics into a  model of enjoyment in games, GameFlow, that is based on flow model and adds
a new ``social interaction'' element. Elmqvist et al. defined \emph{fluid interaction} in the context of information visualization: ``Fluidity in visualization is a concept characterized by smooth, seamless and powerful interaction; responsive, interactive and rapidly updated graphics; and careful, conscientious, and comprehensive user experiences.''~\cite{Elmqvist_flow}
A fluid information visualization interface has three properties: it promotes flow, supports direct manipulation, and minimizes gulfs of action. Although Elmqvist et al.~\cite{Elmqvist_flow} suggest that {\em interactions} should be designed to promote flow, they do not discuss how each of the elements of the flow model can be applied to information visualization in general. They also do not describe how to map the flow elements to visualization tasks or how to measure each of the elements.

\section{Adapting Flow for Information Visualization}
In this section, we discuss how each individual element of flow model can be linked to InfoVis, how these elements correspond to levels of Munzner's model, and how previous work has indirectly measured some of these flow elements.

  \paragraph*{Challenge:} Generally speaking, enjoyment occurs when the challenge in an activity matches the skills of the participant.~\cite{Csikszentmihalyi}. For example, Alper et al.~\cite{IV_Alper2} compared node-link diagrams with matrix representations to assess which representation best supports weighted graph comparison tasks. They showed that participants who were not familiar with the matrix representation of graphs had more difficulty performing the tasks than when the graphs were represented by node-link diagrams. Such results support the idea that challenges in a visualization should match the skills of participants.
In terms of explicitly making visualizations more or less challenging, the work of Hullman about visual difficulties is recently the best-known~\cite{Hullman_Engagement}, in connection to Bateman et al.'s study on visual embellishments, and subsequent followups by Li et al., Borgo et al. and Ghani and Elmqvist.~\cite{CHI_SB, IV_RB, IV_SG, IV_li2014}. We defer a deeper discussion of this point to Section~\ref{sec:discussion}.
 
  \paragraph*{Focus:} Enjoyable activities require complete attention on the task at the hand.
Visualization design has a significant perceptual impact. As advocated by Tufte, the visualization should make it possible to concentrate on the important information~\cite{Tufte1}. This aspect of enjoyment is broken down below into ``clear goals'' and ``immediate feedback'', both of which make very good sense in the context of visualization. In our literature search, we were unable to find papers that specifically discuss user focus during visualization evaluation; feedback and goals, on the other hand, are widely discussed.

  \paragraph*{Clarity:} Enjoyment occurs because the user understands exactly what the task's goals are, and what they're working towards. 
    The clarity of a goal, perhaps surprisingly, is not directly related to the encoding of a visualization or to the data used to generate it. Instead, it is related to the problem and domain in which the user is working. Clarity on the surface appears to be in contradiction to Challenge, since the clearer a goal is, the ``easier'' it is to achieve it. But in fact, these two concerns operate on entirely different levels, and this becomes clear when we map them to the nested model. Consider, for example, the abstract task of finding a path between two nodes of a graph. Node-link diagrams with a reasonable layout make it clear what to do and how to do it (follow paths from the source to the target); with an adjacency matrix representation, the task is less clear~\cite{Fekete_Evaluation}. But the \emph{goal} is likely to have been described as ``find a connection between the suspect and the convicted felon'', and that's irrespective of the encoding or the data being used. If we were to describe the goal differently, then it seems conceivable that the very same visualization \emph{and task} would have different levels of enjoyment. This is an important point we turn back to in Section~\ref{sec:discussion}.
 
  \paragraph*{Feedback:} Enjoyment happens because the task undertaken provides immediate feedback. Liu and Heer~\cite{Liu:2014:TEO} have studied how latency influence the exploratory behavior of users. They results indicate that increasing latency decreases the user performance and causes users to shift exploration strategy. An intriguing possibility is to study the degree to which these shift in exploration strategy comes from disengagement or frustration. We note that alternate feedback mechanisms, especially when the user's visual field is already occupied, is a difficult and well-discussed topic~\cite{Saket-TalkZones}. Visualization designers should be careful of designing interruptive feedback. In other words, there is evidence that there can be \emph{too much} feedback.

  \paragraph*{Control:} When achieving flow, one experiences a sense of complete control over one's actions. Feedback relates to the immediate acknowledgment of an action having happened. Control, on the other hand, relates to the action doing \emph{what one expected it to}. A visualization system should make it possible to translate intentions into visualization behavior and provide a feeling of control. We highlight here the recent Kondo and Collins's DimpVis, where direct manipulations of the visual marks are translated into the data query that would best generate an output with the manipulated configuration~\cite{Collins_Dimpvis}.
    
  \paragraph*{Immersion:} Participants lose their sense of self and become ``lost'' in the activity. Although immersion is frequently discussed in leisure activities such as gaming~\cite{IV_Johnson, Sweetser_flow}, achieving immersion through multisensorial stimulation has long been a goal of virtual reality systems in scientific visualization~\cite{vanDam}. With the ubiquity and dropping prices of virtual-reality equipment, it would appear to be possible to design visualization tasks than can be completed in immersive and non-immersive systems, and then compare participant reports.

\section{Measuring flow in visualization}
Although there is no single holistic method to measure these elements, several studies applied self-reporting methods (e.g., Likert scale questionnaires and Product Reaction Cards~\cite{Mercun_Engagement}) to measure some of these elements in different studies. 

For example, Sweetser et al.~\cite{Sweetser_flow} used the Likert scale to measure the strength of each individual element of their GameFlow model which is derived from the flow model. They first asked participants to play with 
a game. They then designed a set of criteria for each element(e.g.,``games should include online help so players don't need to exit the game''\cite{Sweetser_flow}). They finally asked participants 
to measure how well the game met these criteria with a rating based on the Likert scale (e.g., from completely disagree to completely agree).

In another study, Mer\v{c}un~\cite{Mercun_Engagement} indicates how the product reaction card method can be applied to extract user comments and thoughts on different visualizations. Mer\v{c}un first asked 
participants to work with a particular visualization. She then asked them to select from a set of cards (adjectives and phrases) those that best reflect their experience/feeling about the visualization system or 
technique. The participants were also asked to comment on their choices, thus extracting commentary and providing a better insight into user experience. 

Other methods beyond the Likert scale and reaction cards, such as the HCI-Q method~\cite{HCI-Q}, might be better at evaluating presence of the elements of flow. While the HCI-Q method might give more accurate results, it is relatively new and not well examined. More established methods, such as the Likert scale, have been successfully applied in many domains.

\paragraph*{Recommendations:} While it remains unclear how to design specific measurement methodologies for enjoyment in visualization, the current best model for enjoyment in psychology has several relatively well-defined aspects. In future studies that evaluate engagement in visualization, then, we recommend authors to elicit responses along these different elements. As we have found in the literature, studies specific to one technique or system have touched various aspects of flow, but in order to paint a more complete picture of what ultimately leads to engagement and enjoyment, we ideally need information on all aspects suggested by Csikszentmihalyi~\cite{Csikszentmihalyi}.

\section{Discussion and Limitations\label{sec:discussion}}

Undoubtedly, there will be difficulties in measuring elements of enjoyment. We want to highlight one potentially important concern in comparing different studies. As it relates to visualization, Clarity comes not from the technique, data, or performance, but rather comprehension of the task being performed. In this sense, in order to compare Clarity across visualizations, it seems essential to precisely control the task performed. However, the task typically comes from the task list created by the experimenter, and this information is rarely published along with the study. This, of course, is similar to the problem of \emph{survey question wording}~\cite{Schuman:1981:QAA}.
Does a difference in enjoyment ultimately arise from the relative difference in Clarity between the tasks? This confounding factor seems to require a change in how we report our studies.

While evidence suggests that optimal enjoyment occurs with moderately challenging activities~\cite{LomasGaming,Csikszentmihalyi2} and moderate feedback~\cite{Sweetser_flow,LomasGaming}, we do not have enough 
evidence to draw conclusions about other elements of the flow model.
We illustrate the current situation in Table~\ref{tab:intensity}.

\begin{table}[t]
  \centering
  \includegraphics[width=\linewidth]{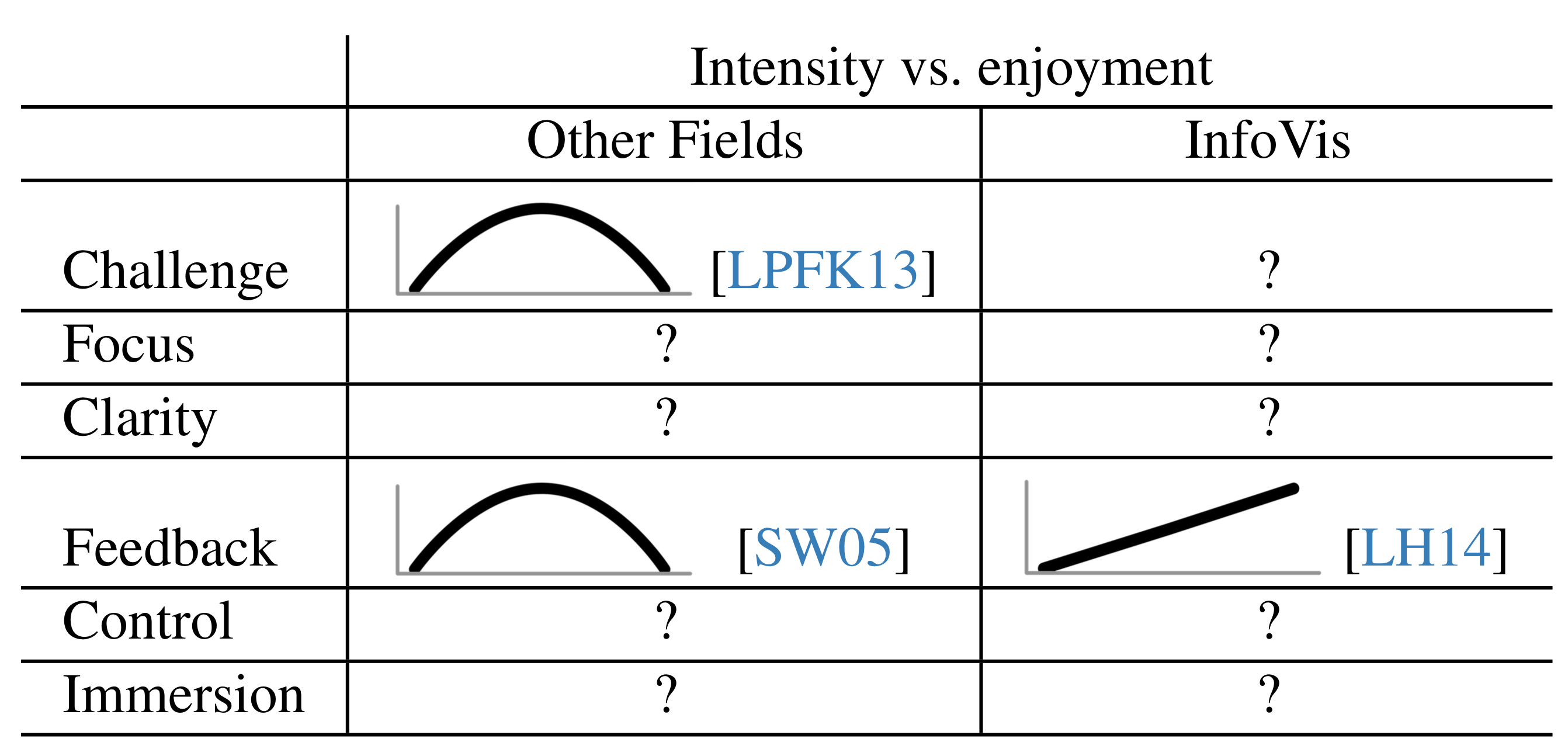}
 
  \caption{For some specific elements of flow, there is evidence that optimal enjoyment occurs with \emph{moderate} intensity levels. This appears to be in contradiction to some published work in visualization~\cite{Liu:2014:TEO}. Question marks indicate areas authors were unable to find published work.\label{tab:intensity}}
\end{table}

\section{Conclusion and Future Work}
In this paper, we connected Csikszentmihalyi's flow model of enjoyment to Munzner's nested model of visualization evaluation.
Regardless of the many hurdles mentioned above, we believe it is important to explore different elements of enjoyment in the context of visualization. 
Our main recommendation is that since ``enjoyment'' encompasses many different elements, in order for visualization researchers to initiate a systematic study of enjoyment in visualization, evaluations must control as many specific flow elements as possible: Challenge, Focus, Clarity, Feedback, Control, Immersion.
We are planning to study enjoyment in the context of node-link and map-based visualizations and will study each flow element specifically; nevertheless, we hope other readers will find our discussion, guidelines, and especially the myriad unresolved research questions, relevant and interesting.

\bibliographystyle{plain}
\bibliography{Bibliography}

\end{document}